
\magnification=\magstep1
\baselineskip 22 pt

\font\temmp=cmssdc10 at 15pt
\temmp
\centerline {Controlling Chaos using an Exponential Control}

\rm
\bigskip

\centerline {Sangeeta D. Gadre and V. S. Varma}

\centerline {Department of Physics and Astrophysics}

\centerline {University of Delhi, Delhi, India.}

\bigskip
\bigskip
\noindent
\bf {Abstract}
\rm

We demonstrate that  chaos can be controlled using a multiplicative exponential
feedback control. All three types of unstable orbits - unstable fixed points,
limit cycles and  chaotic trajectories can be stabilized using this control.
The control is effective both
 for maps and  flows. The control is significant,
particularly for  systems with several degrees of freedom, as  knowledge
of only one variable on the desired unstable orbit is sufficient to settle
the system on to that orbit. We find, that in all the cases studied, the
transient time
is a decreasing function of the stiffness of control. But increasing
the stiffness beyond an optimum value can increase the transient time.
 The control can also be used  to create suitable new stable attractors in a
map, which did not exist in the original system.
\bigskip
\bigskip

\vfil
\eject

\noindent {\bf 1 \quad INTRODUCTION}
\rm

The problem of controlling chaos has recently received much attention [1-8]. A
chaotic system in general cannot be made to converge to a freely evolving
desired trajectory, whether periodic or chaotic, because of the inherent
unpredictibility of the system. The control of chaos in this context consists
of forcing the system to evolve along a desired trajectory. Pecora and Carroll
[1-3] demonstrated that two identical chaotic systems driven by a common signal
display asymptotic convergence of their trajectories even though they may
 have started from very different initial conditions provided the Lyapunov
exponents of the driven system are all negative. Such
 behaviour was  demonstrated numerically and also confirmed experimentally.
\smallskip
Ott, Grebogi and Yorke (OGY) [4,5] succeeded in forcing a chaotic system on to
one of its own unstable periodic orbits by making a set of small time-dependent
perturbations on the system parameters. They demonstrated their method
numerically by controlling
the H\'enon map. They also pointed out that as an infinite number of unstable
periodic orbits are embedded in a chaotic attractor, the stabilization of these
unstable periodic orbits can lead to an enhancement of the system's performance
as well as to the adaptibility of the system to varying performance
requirements since the system behaviour can be changed by  perturbing the
system to stabilize different unstable orbits.  In the absence of chaos,
separate systems would be required for each of the different responses. The
effectiveness of the OGY method of control has been confirmed
experimentally in several systems [6-8].
\smallskip
In this paper we have  succeeded in implementing a novel method for controlling
chaos. We have used a multiplicative exponential feedback
control on a parameter of the system, with the argument of the exponential
being
 proportional to the feedback response of the system, i.e. the difference
between the desired value and the actual value of one of the suitably chosen
variables of the system.
 Our method is a combination of variable
  and parameteric control and it can stabilize all  three types
 of unstable orbits -  unstable fixed points, limit cycles and   chaotic
trajectories. It is found to  work effectively both for maps and flows.
\smallskip
Consider a general N-dimensional dynamical system
$$\dot{\overrightarrow{X}} = {\overrightarrow {F}( {\overrightarrow
{X}};\mu;t)} \eqno(1)$$
where ${\overrightarrow{X}}\equiv (X_1,X_2,.....,X_N)$  are variables and
 $\mu\equiv(\mu_1,\mu_2,....,\mu_K)$  are  parameters
whose values determine the nature of the dynamics.  The stabilization of a
desired unstable attractor
or a chaotic trajectory is possible by multiplying a suitably chosen
 parameter, say $\mu_r$, in (1) by an exponential feedback
control involving only one suitably chosen variable, say $X_l$, with
 the form of control being given by $$exp[\epsilon (X_l-X_l^s)]. \eqno(2)$$
Here $X_l$ is the actual value of one of the variables of the system after
applying the control,
$X_l^s$ is the desired value of that variable and $\epsilon$ is the "stiffness"
of the control which can take both positive
and negative values.
\smallskip
The dynamics of the modulated system in the presence of control is given by
$$\dot{\overrightarrow{X}}=\overrightarrow{F}(\overrightarrow{X};\mu_1,\mu_2,
...,\mu_rexp[\epsilon(X_l-X_l^s)],.., \mu_K ; t). \eqno(3)$$
Note that the control becomes passive once the
desired goal $\overrightarrow{X^s}$ is achieved. If fluctuations drive the
system  off the desired goal, the control reactivates.
\smallskip
The control works for those combinations of controlling parameters and
variables of the system for which the largest real part  of the  Lyapunov
exponents of the modulated system, represented by eq. (3), is negative. The
feedback function in the expression of the control involves only one suitably
chosen variable $X_l$
to convert  the desired repellor - whether a fixed point,
a limit cycle or a chaotic orbit, into an attractor. This shows that the
knowledge of only one variable on the desired unstable orbit is sufficient
to settle the system on to that orbit. This makes the control particularly
useful for  systems with several degrees of freedom.
For example, a desired unstable fixed point of the Lorenz system [9]
can be stabilized using  control with feedback depending on any one of the
X, Y or Z variables whereas an unstable limit cycle or a chaotic trajectory
of the Lorenz system can be converted into an attractor using  feedback
depending on  the Z variable only. For stabilizing most repellors,
it is sufficient to multiply one chosen parameter by the exponential
control; but sometimes, as in the case of one of
the two fixed points of the H\'enon map [10], both the parameters a and b of
the system
have to be multiplied by the exponential control.  Although in all the cases we
have studied,  only one variable on the desired orbit is sufficient to  make
all the Lyapunov exponents of the system  negative, it may not be possible in
some other systems where more than one variable may be required for the
stabilization of unstable orbits. But the fact remains that the control uses
only a subset of the variables and the parameters for controlling chaos.  We
have tested our control for stabilizing different types of
orbits for the logistic map, the H\'enon map and the Lorenz system and found it
to  work effectively.
\smallskip
A quantity of obvious interest in the context of controlling chaos
is the time required for the system to settle on to the desired orbit.
This of course depends upon the stiffness of the control i.e. $\epsilon$.
 For a given $\epsilon$ we study the time $\tau$ required for the system
to approach within a distance $\omega$ of the desired orbit starting
from some initial point. If $\omega_o$ is the initial distance from the
desired orbit, then it is clear that the length of the transient $\tau$
and $\omega$ are related by $\omega = \omega_o exp(\lambda\tau)$, where
$\lambda$ is the largest real part of the  Lyapunov exponents. The slope of the
plot of $\tau$ against $\ln (\omega/\omega_o)$ is nothing but $1/\lambda$. This
of course has to be negative for convergence.
 The values of the Lyapunov exponents calculated from such points  are found to
be in good agreement
with those obtained  either numerically  using the method given in [11] or
analytically  wherever this is  possible. We have studied the transient time
$\tau$ required for settling on to a desired orbit
 within  a given accuracy $\omega$ as a function of $\epsilon$ and found that
$\tau$ in general is a decreasing function of $\epsilon$. But there exists an
optimum stiffness of control beyond which increase in $\epsilon$  increases
$\tau$. This  behaviour of $\tau$ is found to be the same as that of the
$\lambda$ with $\epsilon$ for any orbit.
\smallskip
Finally, for discrete maps we have also tried to stabilize  orbits  which are
not the natural fixed points (stable or unstable) of the original system. We
have found that the control succeeds in creating  desired new stable attractors
 which are not the natural attractors of the unmodulated system, depending on
our requirement and use, using the discrete map. However we cannot stabilize
any arbitrary orbit. The functional form of the map and the control criterion
decides which orbits can be forced on to the system.
\smallskip
There is one drawback. The form of the control is such that for
a given system the unstable fixed point represented by a null vector cannot
be stabilised. The reason is  obvious. The Jacobian
of the modulated system given by (3) evaluated at such a point in the presence
of the control is the same as that of the unmodulated system (1). So the
eigenvalues
of the system remain unchanged in the presence of the control. Hence such
a fixed point remains unstable even under control. Such points may, however,
get
stabilized in the process of stabilizing other fixed points, resulting in the
coexistence of more than one attractor.
\smallskip
The organization of this paper is as follows. In section 2 we study
exponential control for stabilizing an unstable fixed point. We deal first with
1-D discrete systems, then with 2-D discrete systems and finally with
continuous dynamical systems. In section 3.1 we extend the control algorithm so
as to stabilize unstable closed orbits for flows. We suggest a more effective
control for stabilization of higher orbits of a discrete dynamical system  in
section 3.2. In section 4 we discuss the stabilzation of chaotic orbits. We use
our control for stabilization of arbitrary fixed points for discrete systems in
section 5. Finally we summarize our results in section 6.

\bigskip
 \noindent \bf {2 \quad EXPONENTIAL CONTROL FOR STABILIZING AN \hfil \break
UNSTABLE FIXED POINT}

\bigskip
 \noindent \rm{ 2.1\quad 1-D Discrete System}
\rm

\qquad\qquad A representative one dimensional map is the logistic map given by
$$X_{n+1} = 4\mu X_n
(1-X_n) \equiv F(\mu, X_n) \eqno(4)$$ where $0 \leq X$, $ \mu \leq 1$. The map
has two period one fixed points corresponding
to any value of $\mu$, which are ${X^*_a} = 0$ and ${X^*_b} = 1- 1/(4\mu)$. The
fixed point ${X^*_a}$ is unstable for $\mu > 0.25$ while
 ${X^*_b}$ is unstable for both $0 \leq  \mu < 0.25$ and
$\mu > 0.75$. As mentioned in the introduction, the control is ineffective
in stabilizing the unstable fixed point ${X^*_a}$.
\smallskip
For stabilizing  ${X^*_b}$, we multiply $\mu$ by
$exp[\epsilon (X_n - X^*_b)]$ in (4) so that the logistic map in the
presence of control is given by $$X_{n+1} = 4\mu exp[\epsilon (X_n - X^*_b)]
X_n
(1-X_n) \equiv \cal F \rm(\mu, X_n). \eqno(5)$$
 For a
given $\mu$ and for $\epsilon$ in the range $\epsilon_{min} < \epsilon <
\epsilon_{max}$,  the Lyapunov exponent is negative (i.e. $ {\rm d \cal{F}
\over \rm dX}\!\!\!\mid_{_{X^*}}$ lies in the interval $(-1,1)$)
and hence $X^*_b$ becomes a stable fixed point (an
attractor) of the modulated map.  (This is found to be a general feature of the
control
for  maps as well as for flows.) The actual expressions for $\epsilon_{min}$
and
$\epsilon_{max}$ for (5) are: $\epsilon_{min} = 4\mu(4\mu-3)/(4\mu-1)$ and
$\epsilon_{max} =4\mu$.
\smallskip
For a given $\mu$ and $\omega$, the transient  $\tau$ is found to be the
decreasing function of $\epsilon$, however, there is an optimum stiffness
control, beyond which increasing $\epsilon$ increases the transient time. This
behaviour is also  seen in the variation  of the $\lambda$ with $\epsilon$.
The optimum stiffness of control  corresponds to that value of $\epsilon$
for which $\lambda$ is minimum. For the logistic map, $\lambda = -\infty$ for
$\epsilon = 4\mu(4\mu-2)/(4\mu-1)$ and the system settles to $X^*_b$ more
slowly
both for larger and smaller $\epsilon$.

\bigskip
\noindent \rm {2.2\quad 2-D Discrete System}
\rm

\qquad\qquad We have studied the H\'enon map [10] as an example of 2-D maps. It
is
given by $$\eqalign {X_{n+1}& = {Y_{n} + 1 -a X_{n}^2}\cr
Y_{n+1}& = {b X_{n}}\cr}\eqno(6)$$ where
X and Y are variables and a and b are the controlling parameters. The map has
two critical points $(X_{\pm}, Y_{\pm})$ where $$\eqalign {X_{\pm}&=
{(2a)^{-1}(-1+b \pm \sqrt{(1-b)^2+4a} )}\cr
Y_{\pm}& = {bX_{\pm}}\cr}.$$ These two critical
 points are unstable for a = 1.4 and b=0.3.
\smallskip
To stabilize $(X_+,Y_+)$, we multiply the parameter b by
$exp[\epsilon(X_n-X_+)]$ so that the dynamics in the presence of control is
given by
 $$\eqalign{X_{n+1}& = {Y_{n} + 1 -a X_{n}^2}\cr
 Y_{n+1}& = {b exp[\epsilon(X_{n}-X_+)] X_{n}}.\cr} \eqno(7)$$

\smallskip
The other unstable critical point becomes stable only when both parameters a
and b are multiplied by $exp[\epsilon(X_{n}-X_-)]$, so that the map in the
presence of the control for
stabilizing $(X_-,Y_-)$ is given by $$\eqalign {X_{n+1}& = {Y_{n} +
1-aexp[\epsilon(X_{n}-X_-)] X_{n}^2}\cr
Y_{n+1}& = {bexp[\epsilon(X_{n}-X_-)] X_{n}}.\cr}\eqno(8)$$  For a given value
of a and b, the behaviour of the transient time $\tau$  with $\epsilon$ for a
given accuracy $\omega$ is similar to that seen  in discrete
1-D maps.

\bigskip
\noindent \rm { 2.3\quad Continuous Dynamical System}
\rm
\medskip

\qquad\qquad We have studied the Lorenz system [9] as an example of a
continuous
dynamical system. This is governed by the  equations $$\eqalign{\dot{X}& =
{\sigma(Y - X)}\cr
\dot{Y}& = {-XZ + rX - Y}\cr
\dot{Z}&= {XY - bZ}\cr}\eqno(9)$$
where X, Y, Z are variables and $\sigma$, r, b are the controlling parameters.
The system has three critical points, \it{viz} \rm $$X' = 0, Y' = 0, Z' = 0$$
$$X'' = \sqrt{b(r-1)}, Y'' = \sqrt{b(r-1)}, Z'' = r-1$$ and $$X''' =
-\sqrt{b(r-1)}, Y''' = -\sqrt{b(r-1)}, Z''' = r-1.$$ We choose $\sigma = 10$,
$b =8/3$ and
$r = 60$ for which all  three critical points are  unstable fixed points of the
Lorenz system. The first critical point $(X',Y',Z')$ cannot be made  stable
using our form of control. This has been discussed in the introduction.  The
second and third critical points can be stabilized by multiplying  b by
$exp[\epsilon(Z-Z'')]$ and  $exp[\epsilon(Z-Z''')]$ respectively. It is also
possible to stabilize these points using other combinations of  parameters and
variables, provided the control criterion i.e., the largest real part of the
Lyapunov exponents, $\lambda$, is negative,
 is satisfied. Such combinations, with the corresponding values of $\lambda$,
are listed
in table I for $\epsilon = -0.001, -0.01$ and  $-0.1$. Similarly,  $\lambda$
can be worked out for  positive values of $\epsilon$ also. The table shows that
 the r-X, r-Y and b-Y forms of  control work for suitably chosen  negative
values of $\epsilon$. On the other hand, it is the r-Z, b-X and b-Z forms of
control which are effective for suitably chosen positive values of $\epsilon$.
For the rest of the forms \it{viz.} \rm $\sigma$-X,  $\sigma$-Y and
$\sigma$-Z, the value of $\lambda$ remains positive for all  $\epsilon$ (the
Lyapunov exponents  with these forms of  control are independent of $\epsilon$
and are the same as that of the unmodulated system) thereby implying that these
forms of exponential  control cannot stabilize the unstable fixed points of the
Lorenz system.
\smallskip

For a fixed $\omega$, the  transient time $\tau$,  using b-Z exponential
control, is found to be a decreasing function of  $\epsilon$. However there
exists an optimum value of $\epsilon$ beyond which increasing $\epsilon$
increases $\tau$ (as in the case of discrete 1-D maps). The Lyapunov exponents
obtained analytically for the second and third critical points in the presence
of  control have the same value. We plot $\lambda$ vs $\epsilon$ in fig. (1).
The variation of $\lambda$ with $\epsilon$ shows the same behaviour as that of
$\tau$ with $\epsilon$. The  optimum value of $\epsilon$ for which $\tau$ is
minimum also corresponds to the most negative value of $\lambda$ and is $0.2$
for the given values of the parameters. The value of $\lambda$  obtained
analytically is $-6.8993$ for $\epsilon = 0.2$.  On plotting  $\tau$, as a
function of $\ln (\omega/\omega_o)$ for the same value of $\epsilon$, we find
that the reciprocal of the slopes for the second and third critical po!
 ints are $\approx -6.92$,
which is in good agreement with the value obtained analytically.
To compare our results  with those of  Pecora and Carroll, with both X and Y
drives, we first consider the control with the  parameter r  multiplied by the
exponential feedback function involving the X variable for the same parameter
values (such a choice of the control is made because Pecora and Carroll's
method uses variables X and Y as drives). We find that the value of $\lambda$
with this form of  control is $-4.536$  for $\epsilon = -0.17$. The value of
$\lambda$ for  control involving the parameter r and the variable Y  is
$-7.593$  for $\epsilon = -0.05$, for the  same parameter values. The
corresponding values of
$\lambda$ in  the method suggested by Pecora and Carroll for the same parameter
values and are $-1.83$ for the X drive and $-2.85$ for the Y drive [12]. Thus
our values of $\lambda$ are more negative and hence the control asserts itself
quicker. This can be seen in fig. (2) for the case of  the second unstable
fixed point of the Lorenz system.  This result is not surprising in view of the
fact that our control has an exponential form.

\bigskip
\bf\noindent {3 \quad  EXPONENTIAL CONTROL FOR STABILIZING  \hfil \break
UNSTABLE HIGHER PERIOD ORBITS}

\bigskip
 \noindent \rm{ 3.1\quad  STABILIZATION OF UNSTABLE CLOSED ORBITS \hfil \break
FOR FLOWS}
\rm

The unstable limit cycle of a continuous system  can be converted
into a stable attractor by multiplying one  suitably chosen
parameter, say $\mu_l$, (or may be more parameters depending on the system) by
the exponential
 feedback function which depends only on one of the variable say $X_r$,
i.e. $$\mu_l \rightarrow \mu_l exp[\epsilon(X_r-X_r^u)]$$ where $X_r$ is
 the actual value of the variable of the given system with the feedback
 control under consideration and $X_r^u$ is the value of the $X_r$ coor
dinate on the desired unstable limit cycle which is required to be stabilized.
 $X_r^u$ is obtained by allowing the system to evolve freely on the desired
unstable orbit with  the same parameter values but without imposing the
control.
\smallskip
We have implemented this idea for controlling unstable limit cycles of the
Lorenz system. We choose parameter values $\sigma=10$, $b=8/3$, $r=28$.
For these parameters one of the unstable limit cycles has a point with
coordinates ($-12.786189$, $-19.364189$, $24.00$) [13]. The $\lambda$ of the
system is found to be positive. Since the limit cycle
is unstable, so starting with an initial state, even very close to
the limit cycle, the trajectory will diverge away from the limit cycle.
 But after implementation of the control, multiplying parameter b by
$exp[\epsilon(Z-Z^u)]$, the Lyapunov exponents are found to be  functions of
$\epsilon$ and  for a certain range of values of $\epsilon$, the real part of
all the Lyapunov exponents become negative  thereby implying the stabilization
of the limit cycle. Now, nearby trajectories are found to converge to the
desired limit cycle as shown in fig. (3) also. The basin of attraction of the
limit cycle is not infinite implying that there is a set of  initial values
starting from which the trajectories converge to the desired limit cycle.
Otherwise they  may escape to infinity. The transient $\tau$ shows a similar
trend as in the previous sections. But for a given accuracy $\omega$
the minimum value of $\tau$ corresponding to the optimum $\epsilon$ is much
higher  compared to the case of the unstable fixed point of the Lorenz system.
Moreover, the variation of $\tau$ with $\epsilon$ for a given $\omega$
is not as rapid as in the case of the unstable fixed point. Consequently
the $\tau$ vs $\epsilon$ plot is quite flat and an optimum $\epsilon$ cannot
be obtained very accurately.
\smallskip
We have also tested the control for another limit cycle of the Lorenz
attractor. For the same parameter values, the coordinates of a point on a
different limit cycle are $(-13.917865,$ $-21.919412,$ $24.00)$ [13]. The
control was found to work effectively in this case with the same features as
have been listed above.

\bigskip
\noindent\bf{3.2\quad STABILIZATION OF HIGHER PERIOD ORBITS OF A \hfil \break
DISCRETE DYNAMICAL SYSTEM}
\rm

\qquad\qquad We can extend the above control to convert unstable fixed points
of higher period say $(\overrightarrow {X^{*1}}, \overrightarrow
{X^{*2}},......,\overrightarrow {X^{*k}})$ to  stable fixed points for given
values of the control parameters. What is required is a feedback that encodes
as much information about the periodic orbits as is necessary for its unique
characterization. But there is a practical problem here. Since in this case the
form of the control given by eq. (2) requires the convergence of the chosen
variable to  a set of k values of that variable, the controlling technique
diminishes in utility with increase in period. For higher period orbits of a
discrete dynamical system the more effective control is  one which employs a
logical OR
structure in the feedback function. So in order to stabilize  unstable fixed
points of period k \it {i.e.} \rm $\{\overrightarrow {X^{*i}}\}$ one of the
parameters say $\mu_{j}$ in (1) is multiplied by $$exp[\epsilon \prod_{i=1}^k
({(X_l)}_n-X_l^{*i})] \eqno(10)$$ where ${(X_l)}_n$ is the value of one of the
chosen variables $X_l$ at time $= n$ of the modulated map after applying the
control and $({X_l}^{*1}, {X_l}^{*2},....,{X_l}^{*k})$ is the set of k values
of the variable $X_l$ on the desired unstable period k orbit.

We implemented this in the case of  the logistic map for stabilizing period 2
orbit
 having $X^{*1}$ and $X^{*2}$ as the fixed points and found it to  work
effectively.

\bigskip
\noindent\bf {4 \quad STABILIZATION OF CHAOTIC ORBIT}
\rm

\qquad\qquad\rm We have also tried to converge different trajectories to a
particular desired chaotic trajectory using our control. Again taking the
Lorenz system as an example we choose parameter values  $\sigma = 16$, $b = 4$,
$r = 40$. The real parts of the Lyapunov exponents  are ($1.37$, $0.0$,
$-22.37$~) [14] implying that this is a chaotic
regime. We  chose a chaotic trajectory starting with  initial
 coordinates ($10.0$, $0.0$, $30.0$). We applied our control by multiplying b
by $exp[\epsilon(Z-Z^c)]$, where
$Z^c$ is the Z coordinate of a point on the freely evolving chaotic trajectory.
We  found that the real parts of all the Lyapunov exponents of the modified
Lorenz system
become negative for a certain range of values of $\epsilon$, showing that the
desired chaotic trajectory has become a
stable trajectory and different closeby trajectories starting from different
initial states converge to the desired trajectory. For a given $\epsilon$ the
reciprocal of the
slope of $\tau$ vs $\ln (\omega/\omega_o)$ is in good agreement with the
minimum value of $\lambda$. It has been verified
numerically that the system settles down on to the desired trajectory in the
presence of the control. The features of the control remain the same as in the
previous section. We have also tested our control for  other chaotic
trajectories and found it to work effectively.

\bigskip
\noindent\bf {5 \quad  STABILIZATION OF ARBITRARY FIXED POINTS FOR \hfil \break
DISCRETE SYSTEM}

\rm
\qquad\qquad\rm With exponential control it is possible to create new stable
attractors which do not exist in the unmodulated system. This allows the
modulated system to settle down to arbitrary fixed points which are not the
fixed points (stable or unstable) of the unmodulated map. As an example we
again  take the
logistic map given by eq. (4). Suppose the requirement is to stabilize the
system to say a period 2 attractor $(X_1, X_2)$ which does not exist in the
unmodulated map, implying that $X_1$ and $X_2$ are not the fixed points of
$F^2(\mu ,X)$. The equations governing the dynamics in the presence of the
control are $$\eqalign{X_{n+1}& ={4\mu_1 exp[\epsilon (X_n - X_1)(X_n - X_2)]
X_n(1 -X_n)}\cr
\rm X_{n+2}& = {4\mu_2 exp[\epsilon (X_{n+1} - X_1)(X_{n+1} - X_2)]
 X_{n+1}(1 -X_{n+1})}\cr} \eqno(11)$$ where $\mu_1$ and $\mu2$ are given by
$$\mu_1= X_2/(4X_1(1-X_1))$$  and $$\mu_2 = X_1/(4 X_2(1-X_2))$$ Any arbitrary
combination of $X_1$ and $X_2$ cannot be stabilzed using eq. (11). Only those
combinations of $X_1$ and $X_2$  can be stabilized in the logistic map using
the  exponential control for which the $\lambda$  of the modulated map is
negative. The coordinates of the points lying in the region enclosed by the
curves and the coordinate axes in fig. (4) represent such combinations of $X_1$
and $X_2$ for which $\lambda$ is negative, and only such combinations can be
stabilized using the map given by eq. (11).
\smallskip
This procedure can easily be extended to stabilize the system to higher period
orbits even when the points on the orbit do not correspond to either
stable or unstable fixed points of the higher period orbits of the original
system.
\bigskip
\noindent
\bf
6 \quad CONCLUSION

\rm  Exponential control is found to be  effective
both for maps as well as for flows. Since the knowledge of only a subset of
variables on the desired unstable orbit is sufficient to settle the system on
to that
orbit, this makes the control more significant particularly for
systems with several degrees of freedom.
 In the presence of  control, there
exists a range of values of the stiffness constant $\epsilon$  for which a
repellor - an unstable fixed point, a limit cycle or a chaotic trajectory, of a
system
can be converted into an attractor. The $\lambda$
is found to be the function of $\epsilon$. The  transient time $\tau$ is a
decreasing function of $\epsilon$  for a given accuracy $\omega$ . However
there exists an
 optimum stiffness control beyond which increasing
$\epsilon$ can increase  $\tau$. This behaviour of $\tau$ is found to
hold  for different types of orbits of different systems and is similar to the
behaviour  of  $\lambda$ with $\epsilon$. The optimum $\epsilon$  corresponds
to the value of $\epsilon$ where $\lambda$ is minimum.
\smallskip
 Exponential control is found to settle the Lorenz system on to  its non-zero
unstable fixed  points faster than the control of Pecora and Carroll.  In
addition, it is
also possible to stabilize arbitrary fixed points, chosen according to one's
requirements,  which do not correspond to
either stable or unstable fixed points of the unmodulated map.
\smallskip
We are  investigating the effect of exponential control in the presence of
noise. It would be interesting to see whether such control is capable of
controlling spatio-temporal chaos in coupled map lattice systems [15]. It would
also be interesting to see  whether  exponential control can successfully
recover the original signal when the original signal has been masked by the
addition of noise [16].
\bigskip
\noindent
\bf
Acknowledgements:

\rm We would like to thank Dr. Neelima M. Gupte for her invaluable comments
and suggestions. We also thank the Inter University Center for Astronomy and
Astrophysics (IUCAA) for allowing the use of their computing facility. One of
the authors (SDG) also thanks Kirori Mal College, University of Delhi, Delhi,
for the grant of study leave.
\vfil
\eject

\noindent {\bf REFERENCES}
\rm

\noindent
\item { [1]} L. M. Pecora and T. L. Carroll, Phys. Rev. Lett. \bf {64}\rm, 821
(1990).

\item {[2]}  L. M. Pecora and T. L. Carroll, Phys. Rev. Lett. \bf {67}\rm, 645
(1991).

\item{[3]}  L. M. Pecora and T. L. Carroll, Phys. Rev. A \bf {44}\rm, 2374
(1991).

\item {[4]} E. Ott, C. Grebogi and J. A. Yorke, Phys. Rev. Lett. \bf {64}\rm,
1196 (1990).

\item {[5]} T. Shinbrot, E. Ott, C. Grebogi and J. A. Yorke, Phys. Rev. Lett.
\bf {65}\rm, 3215 (1990).

\item {[6]} W. L. Ditto, S. N. Rauseo and M. L. Spano, Phys. Rev. Lett.
\bf{65}\rm, 3211, (1990).

\item {[7]} E. R. Hunt,  Phys. Rev. Lett. \bf {67}\rm, 1953 (1991).

\item {[8]} C. Reyl, L. Flepp, R. Badii and E. Brun Phys. Rev. E \bf {47}\rm,
267 (1993).

\item {[9]} E. N. Lorenz, J. Atoms. Sci. \bf{20}\rm, 130 (1976).

\item {[10]} M. H\'enon, Commun. Math. Phys. \bf{50}\rm, 69 (1976).

\item {[11]} C. Sparrow, \it {The Lorenz equations, Bifurcations and Chaos} \rm
(Springer Verlag, New York, 1982).

\item {[12]} N. Gupte and R. E. Amritkar, Phys. Rev. E \bf{48}\rm, R1620
 (1993).

\item {[13]} J. H. Curry, in \it{Global Theory of Dynamical Systems}\rm, Lect.
Notes in Mathematics, vol. 819, Ed. Z. Nitecki and C. Robinson (Springer
Verlag, Berlin Heidelberg, 1980) p. 111-120.

\item {[14]} I. Shimada and T. Nagashima, Prog. Theor. Phys. \bf{69}\rm, 1605
(1979).

\item {[15]} G. Hu and Z. Qu, Phys. Rev. Lett. \bf {72}\rm, 68 (1994).

\item {[16]} K. M. Cuomo and A. V. Oppenheim, Phys. Rev. Lett. \bf {71}\rm, 65
(1994).

\vfil
\eject

\noindent {\bf FIGURE CAPTIONS}
\bigskip
\rm

\noindent Fig.1  The  variation of  $\lambda$ with $\epsilon$ for the Lorenz
system while stabilizing the unstable fixed point $(\sqrt{b(r-1)},
\sqrt{b(r-1)}, r-1)$ using the exponential control involving the parameter b
and the variable Z, with $\sigma = 10$, $b =8/3$ and $r = 60$.
\smallskip

\noindent Fig.2 Plot of the transient time $\tau$ vs $\ln(\omega / \omega_o)$
 for the Lorenz system with  $\sigma = 10$, $b =8/3$ and $r = 60$ while
stabilizing  the unstable fixed point $(\sqrt{b(r-1)}, \sqrt{b(r-1)}, r-1)$
using Pecora and Carroll's method with $X=\sqrt{b(r-1)}$  drive (triangles),
with  $Y=\sqrt{b(r-1)}$ drive (squares), and also with  exponential control
involving the parameter r and the variable X (circles), and  r and Y (stars).
\smallskip

\noindent  Fig.3 The dotted trajectory shows the convergence of a nearby
 trajectory to the unstable closed orbit, represented by the bold curve,
 projected on to the (X,Z) plane of the Lorenz system with $\sigma=10$,
 $b=8/3$, and $r=28$ and a point with coordinates ($-12.786189$, $-19.364189$,
$24.00$).
\smallskip

\noindent Fig.4 The coordinates of the points lying in the region enclosed by
 the plotted curves and the coordinate axes represent the combinations of $X_1$
and $X_2$ which can be stabilized in the logistic map using exponential control
for $\epsilon=0.1$ and $1.0$.

\vfil
\eject

\noindent {\bf TABLE CAPTION}
\rm

\noindent  The values of $\lambda$, the largest real part of the Lyapunov
exponents, of the  Lorenz attractor under exponential control while stabilizing
the fixed point  $(\sqrt{b(r-1)},$ $ \sqrt{b(r-1)}, r-1)$  for $\sigma=10$,
$b=8/3$ and $r=60$  for different combinations of parameters and variables.

\vfil
\eject

\newdimen\tempdim
\newdimen\othick	\othick=.4pt
\newdimen\ithick	\ithick=.4pt
\newdimen\spacing	\spacing=9pt
\newdimen\abovehr	\abovehr=6pt
\newdimen\belowhr	\belowhr=8pt
\newdimen\nexttovr	\nexttovr=8pt

\def\r{\hfil&\omit\vrsp\vrule width\othick\cr&}
\def\rr{\hfil\down{\abovehr}&\omit\vrsp\vrule width\othick\cr
    \noalign{\hrule height\ithick}\up{\belowhr}&}
\def\up#1{\tempdim=#1\advance\tempdim by1ex
    \vrule height\tempdim width 0pt depth 0pt}
\def\down#1{\vrule height0pt depth#1 width0pt}
\def\large#1#2{\setbox0=\vtop{\hsize#1 \lineskiplimit=0pt \lineskip=1pt
    \baselineskip\spacing \advance\baselineskip by 3pt \noindent
    #2}\tempdim=\dp0\advance\tempdim by\abovehr\box0\down{\tempdim}}

\def\vrsp{\hskip\nexttovr\relax}
\def\toprule#1{\def\startrule{\hrule height#1\relax}}
\toprule{\othick}
\def\nstrut{\vrule height\spacing depth3.5pt width0pt}

\def\preamble#1{\def\startup{#1}}
\preamble{&##}
{\catcode`\!=\active
 \gdef!{\hfil\vrule width0pt\vrsp\vrule width\ithick\relax\vrsp&}}

\def\table #1{\vbox\bgroup \setbox0=\hbox{#1}
    \vbox\bgroup\offinterlineskip \catcode`\!=\active
    \halign\bgroup##\vrule width\othick\vrsp&\span\startup\nstrut\cr
    \noalign{\medskip}
    \noalign{\startrule}\up{\belowhr}&}

\def\caption #1{\down{\abovehr}&\omit\vrsp\vrule width\othick\cr
    \noalign{\hrule height\othick}\egroup\egroup \setbox1=\lastbox
    \tempdim=\wd1 \hbox to\tempdim{\hfil \box0 \hfil} \box1 \smallskip
    \hbox to\tempdim{\advance\tempdim by-20pt\hfil\vbox{\hsize\tempdim
    \noindent #1}\hfil}\egroup}

\magnification=\magstep1
\baselineskip 22 pt
\font\ssdb = cmssdc10 scaled\magstep1

\font\temmp=cmssdc10 at 20pt

\table {\ssdb {TABLE 1}} \rm
! \multispan {3}\hfil \rm {$\lambda$  for }\rr \rm
Control!$\epsilon =-0.001$!$\epsilon =-0.01$!$\epsilon =-0.1$\r
\hfil form ! ! ! \rr
r-X!0.616        !$-$0.267      !$-$3.950 \rr

r-Y!0.394	  ! $-$2.381	! $-$7.180 \rr

r-Z!0.767	  ! 1.117	! 2.250 \rr

b-X!0.739	  ! 0.927	! 2.425	\rr

b-Y!0.707	  ! 0.611	! $-$0.113 \rr

b-Z!0.786	  ! 1.403	! 7.884	 \rr

$\sigma$-X!0.717  !0.717	!0.717  \rr

$\sigma$-Y!0.717   !0.717	!0.717  \rr

$\sigma$-Z!0.717   !0.717	!0.717

\caption {\hfil}

\vfil
\eject
\end